\documentclass [12pt]{article}
\usepackage{epsf}

\begin{document}
\begin{center}
\textbf{\Large Quantum mechanics of the closed collapsing
Universe.}
\end{center}

\begin{center}
K. A. Viarenich$^{1}$, V. L. Kalashnikov$^{2}$, S. L.
Cherkas$^{3}$
\end{center}

$^1$ {\it \small Belarusian State University, Nezalezhnasti av.,
4, Minsk  220080, Belarus}

$^2$ {\it \small Institut f\"{u}r Photonik,Technische
Universit\"{a}t Wien, Gusshausstrasse 27/387, $~~~~~~~$Vienna
A-1040, Austria}

$^3$ {\it \small Institute for Nuclear Problems, Bobruiskaya 11,
Minsk 220050, Belarus}

\begin{center}
Abstract
\end{center}

{\small Two approaches to quantization of Freedman's closed Universe
are compared. In the first approach, the Shr\"{o}dinger's norm of
the wave function of Universe is used, and in the second approach,
the Klein-Gordon's norm is used. The second one allows building the
quasi-Heisenberg operators as functions of time and finding their
average values.  It is shown that the average value of the Universe
scale factor oscillates with damping and approaches to some constant
value at the end of the  Universe evolution. }

\bigskip

The unification of quantum mechanics and gravity is one of the most
important problems of modern physics and it is an aspect of the
string theory development [1]. Quantum gravity is necessary for the
description of the early Universe, because it is thought that at the
Planck time $\tau _P = \sqrt {\frac{G\hbar }{c^5}} $ from the Big
Bang, Universe has to be described using quantum mechanics. For the
first time, the canonical formalism including the wave function of
Universe and its configuration space was worked out about 40 years
ago [2,3].

In spite of some success of quantum gravity, there are still many
problems to date. They are absence of time, interpretation of the
wave function of Universe and building of the Hilbert space of
physical states [4-10].

For the closed Universe, the problem of collapse appears. It is
known, that the bodies, which are massive enough, collapse under the
force of gravity and their density grows unlimitedly. The same thing
takes place for the closed Universe filled with matter. According to
common expectations, the quantum description has to solve this
problem, but still there is no answer to this question. On the
contrary, the collapse leads to new difficulties in quantum
description, because it is accompanied by the dissipation of
probability. As it will be seen further, one of the ways to avoid
the problem is admitting of the re-collapse, which means that system
expands again after the collapse.

The solution of two problems will be proposed here: the problem of
time and the problem of collapse.

Einstein's action for gravity and one-component real scalar field
(which is only type of the matter in this model) can be written as
follows:

\begin{equation}
\label{eq1} S = \frac{1}{16\pi G}\int {d^4x} \sqrt { - g} R + \int
d^4x\sqrt {- g} [ \frac{1}{2}( {\partial _\mu \phi
}\,g^{\mu\nu}{\partial _\nu \phi } - V(\phi)) ] ,
\end{equation}

\noindent where $R$ is the scalar curvature [11], $g$=det$\vert
$g$_{\mu \nu }\vert $ is the determinant of co-variant metric,
$V(\phi )$ is the self-acting potential of the scalar field [6],
that may include the cosmological constant.

Let us consider the homogeneous isotropic model of Universe using
metric:

\begin{equation}
\label{eq2} ds^2 = N^2(t)dt^2 - a^2(t)d{\sigma }^2.
\end{equation}

Here $N(t)$ is the lapse function representing the freedom for
transformation of the time coordinate, $a(t)$ is the scale factor of
Universe (the physical distance between any two points grows
according to ${r(t) = r_{0}\,a(t)}$, where ${r_{0}}$ sets some scale
to measure distances).

For the restricted metric (2) of the closed  Universe, the action is
reduced to

\begin{equation}
\label{eq3} S = \int N(t)\left( \frac{3}{8\pi G}(a - \frac{a\,\dot
{a}^2}{N^2(t)}) + \frac{1}{2}a^3\frac{\dot {\phi }^2}{N^2(t)} -
a^3V(\phi) \right) dt.
\end{equation}

This form of action follows from varying on $p_{a}$ and $p_{\phi }$

\begin{equation}
\label{eq4}
S = \int {\{ {p_\phi \dot {\phi } - p_a \dot {a} - N(t)({-\frac {3a}{8\pi G} - \frac{8\pi Gp_a^2 }{12a} +
\frac{p_\phi ^2 }{2a^3} + a^3V(\phi)})} \}} dt,
\end{equation}

\noindent and after varying the momentums   $p_\phi =
\frac{1}{N(t)}\dot {\phi }a^3$ and $p_a = \frac{3}{4\pi N(t)}a\dot
{a}$ have to be substituted back into Eq. (4).

The varying of action (3) or (4) on $N(t)$ makes the Hamiltonian
of the system equal to zero on the classic trajectories of motion
of the system:

\begin{equation}
\label{eq5}
H = - \frac{3a}{8\pi G} - \frac{8\pi Gp_a^2 }{12a} + \frac{p_\phi ^2 }{2a^3}
+ a^3V( \phi) = 0.
\end{equation}

To come to the quantum theory we should change the classic
momentums by the appropriate operators, which satisfy the
commutation relations ${[{\hat {p}_\phi},\hat {\phi}]=-i}$,
${[{\hat {p}_a},\hat {a}]=i}$. This can be made by using ${\hat
{p}_\phi}=-{i\frac {\partial}{{\partial}\,\phi}}$, ${\hat
{p}_a}=-{i\frac {\partial}{{\partial}\,a}}$.

The quantum version the of Hamilton constraint is the Wheeler-DeWitt
equation ${\hat H \psi = 0}$ [2, 3], where $\psi(a,\phi) $ is the
wave function of Universe. As one can see, there are the
non-commutative operators $1/a$ and $p_{a}$ in the Hamiltonian (5).
They result in the operator-ordering problem. At first we would like
to consider the wave function of Universe to be normalized according
to the Shr\"{o}dinger's rule ${\int _{0}^{\infty }\!{da} \int
_{-\infty}^{\infty }d\phi\,\!{\psi^* \psi} }$. In this case, one
should choose the operator-ordering so that the wave function turns
to zero when the scale factor is zero or infinity. Thus the
Wheeler-DeWitt equation is:

\begin{equation}
\label{eq6} \frac {1}{{2}}\left(\frac {1}{{a}} \frac
{\partial^2}{{\partial a^2}}-\frac {1}{{a^3}} \frac
{\partial^2}{{\partial \phi^2}}-a-a^3 V(\phi)\right)\psi (a,\phi)=0.
\end{equation}
Here the Planck units $(4\pi G/3=1)$ are used. Below only the
$V(\phi )=0$ case will be considered, and Eq. (6) is solved by the
variable separation method:

\begin{equation}
\label{eq7} {\psi_k}(a,\phi)=\sqrt {a}K_{\sqrt
{1/16-4\,{k}^{2}/4}}(1/2 \,{a}^{2}){e^{ik\phi}},
\end{equation}
where $K_{\lambda }$(z) is the modified Bessel function [12].
Normalized solution for Eq. (6) appears as a wave packet:

\begin{equation}
\label{eq8}
{\psi} (a, \phi)= C \int _{-\infty}^{\infty }\!{\psi_k}(a, \psi ){c} (k) {dk},
\end{equation}
where $C$ is the normalizing factor.

The function $\psi_k$ tends to
\begin{equation}
\label{eq9} \psi_{k} \approx a^ {1/2-\sqrt {1/4-{k}^{2}/4}}
{e}^{ik\,\phi}
\end{equation}
asymptotically under ${a\to 0}$ and becomes zero at $a = 0$ only if
${k \ne 0}$. This means that for constructing of the wave packets
one has to use $c(k)$ turning to zero at $k = 0$, e.g. ${{c} (k) =
{k}^{2}\,{e}^{-{k}^{2}}}$. As a result we have absolutely static
Universe with some distribution of the scale factor and the scalar
field amplitude. This strange fact inspired a lot of speculations,
because it is known that Universe expands. Certainly, it can be
proven that there is no time evolution  in the Heisenberg picture,
too. For example, let us multiply Eq. (6) by $a$ and consider the
operator ${\hat {\mathcal H} = {\hat {\mathcal H}}^{+}=}$ ${\frac
{1}{{2}}(\frac {{\partial}^{2}}{{\partial\,{a}^{2}}} - \frac
{1}{{a^2}}\,\frac {{\partial}^{2}}{{\partial\,{\phi}^{2}}} -
a^{2})}$ as the Hamiltonian of the system. Although we can write
non-trivial Heisenberg operators ${\hat A (t) = e^{i\,\hat {\mathcal
H}\,t}\,\hat A\,e^{-i\,\hat {\mathcal H}\,t}}$, their average values
do not depend on time.

\begin{eqnarray*}
 < {\psi  \vert \dot {\hat A}(t) \psi>} = i\,< {\psi |
(\hat {{\mathcal H}}\, \hat {A} - \hat {A}\hat {{\mathcal
H}})\,\psi} > = < {\psi | \hat {{\mathcal H}}\,\hat {A}\,\psi} >
\\= \int _{0}^{\infty }\!\int _{-\infty }^{\infty }\!({\hat
{{\mathcal H}}}^+\psi (a,\phi))^*\hat {A}\psi(a,\phi){d\phi}\,{da}
\\= \int _{0}^{\infty }\! \int_{-\infty }^{\infty
}\!({{\mathcal H}{\psi}(a,\phi))^*}A \psi(a,\phi){d\phi}\,{da}  =
0.
\end{eqnarray*}

The essential moment in these formulas is that one can move the
differential operators to the left using integration by parts,
because the wave function is equal to zero on the integration
limits.

Now, let us normalize the wave function according to the Klein -
Gordon rule. This seems to be reasonable, because the Wheeler -
DeWitt equation looks very similar to the Klein - Gordon one. In
this case the natural operator ordering is the Laplacian one:

\begin{equation}
 \left(\frac {1}{2a^2}\frac {\partial}{\partial
a}a\frac {\partial}{\partial a}-\frac {1}{2a^3}\frac
{\partial^2}{\partial \phi^2}+\frac {a}{2}\right)\psi=0,
\end{equation}
so that the solutions have the asymptotic
\begin{equation}
\label{eq10} \psi_{k}(a,\varphi)=a^{\pm i\vert k\vert
}e^{ik\varphi}
\end{equation}
similar to the plane wave. The normalization of the solution
 [4,10] looks like:

\begin{equation}
\label{eq11a} < {\psi}^*|\psi > = ia\int \left(\!{\psi}\,{\frac
{d}{da}}\psi^*-\psi^*\,{\frac
{d}{da}}{\psi}\right)\biggl|_{a=0}{d\phi}=1.
\end{equation}

As it was in the first case, the wave function has to be a wave
packet, but obtained from the negative-frequency partial solution
$\psi_k\sim a^{-i |k|}e^{i k\phi}$
\begin{equation}
\psi(a,\phi)=\int c(k)\psi_k(a,\phi)d \,\phi.
\end{equation}

The wave function is not restricted along the $a$ variable  now,
that is why the absence of time evolution in the Heisenberg picture
can not be proved as it was done above. And it gives us an idea that
a time dependent picture (like the Heisenberg's one) could exist.
Such a picture, consisting  in quantization of the equations of
motion, which is compatible with the normalization (12), has been
proposed in Refs. [13, 14], where the Dirac commutation relations
for the quasi-Heisenberg operators at initial moment of time are
postulated. According to Refs. [13, 14] the operator equations of
motion are:

\begin{figure}[h]
\vspace{-1.5cm}\hspace {0. cm} \epsfxsize =13. cm \epsfbox[100 140
410 350]{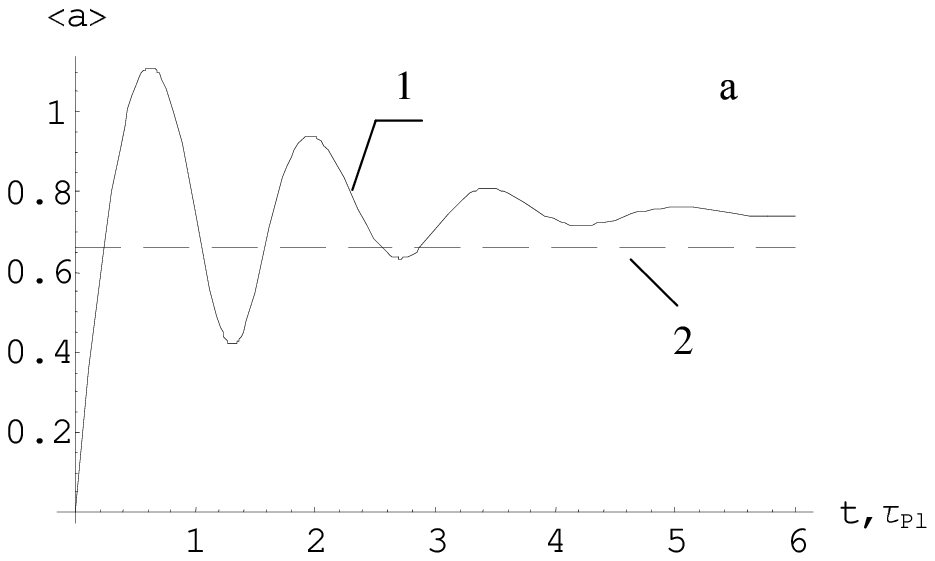} \epsfxsize =14.5 cm \epsfbox[80 140 410
350]{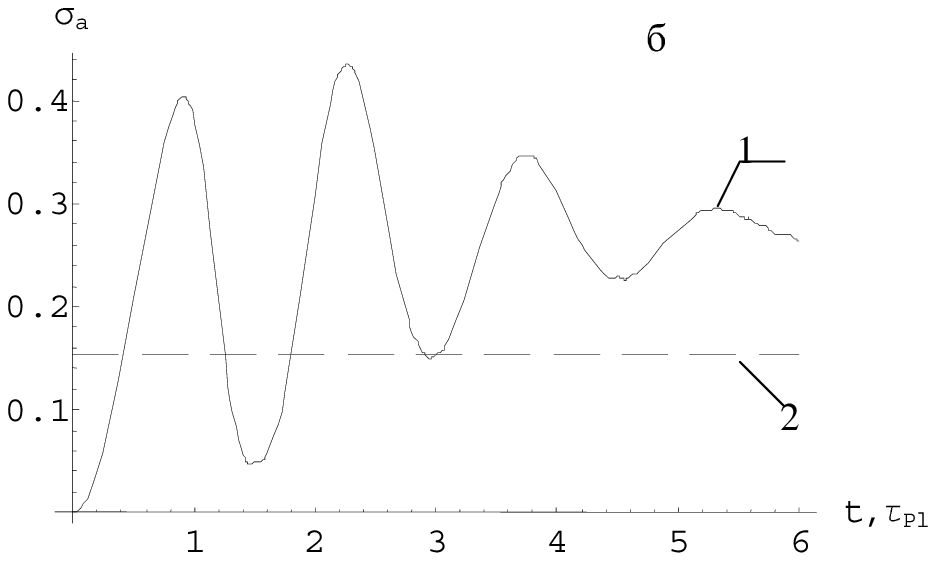} \caption{ Evolution of the mean value of the scale
factor (a) and its dispersion (b) calculated in the
quasi-Heisenberg picture --- (1). Dashed line (2) corresponds to
the fully static picture when the Shr\"{o}dinger norm of the wave
function is used.}
 \vspace{0.5cm}
\label{1}
\end{figure}

\begin{eqnarray}
 \ddot {\hat {a}} = - \frac {\dot {\hat {a}}^2}{2\hat
{a}} - \frac {3 \hat {{p}_{\phi}}^2}{2 \hat {a}^3}-\frac {1}{2
\hat {a}},\nonumber \\
 \dot {\hat {a}} = \frac {\hat {p_a}}{\hat {a}}, ~~~~~~
\dot {\hat {\phi}} = \frac {\hat p_{\phi}}{\hat {a^{3}}},~~~~~
\dot {\hat p_{\phi}} = 0.
\end{eqnarray}

They have to be solved using the following operator initial
conditions:

\begin{equation}
\hat a (0) = const = a,~~ \hat \phi (0) = \phi ,~~ \hat p_{a} (0)
= \frac {\vert \hat p_{\phi} \vert}{a},~~ \hat p_{\phi}(0) = \hat
p_{\phi},
\end{equation}
\noindent where ${\hat p_{\phi}=-i \partial / \partial \phi}$. Thus,
the initial values of operators satisfies the Hamiltonian constraint
\[
-\frac{{\hat p}^{2}_a(0)}{2 \hat a(0)}+\frac{{\hat p}^2_\phi(0)}{2
{\hat a}^3(0)}+\frac{\hat a(0)}{2} =0.
\]
The solution of the equations of motion is found in the momentum
representation most easily. In this representation ${\hat p_{\phi} =
k}$, ${\hat \phi = i \frac {\partial}{\partial k}}$. Finally, if one
sets ${\hat a (0) = const = 0}$, the solution of the equations of
motion (14) are

\begin{eqnarray}
\label{eq18} a(\eta,k)=\sqrt {k}\sqrt [4]{\frac{1-\cos(4\eta)}{2}},\\
 t(\eta,k)={2}^{-9/4}\sqrt {k}\int
_{0}^{4\,\eta}\!\sqrt [4]{1-\cos(\xi)}{d\xi}.
\end{eqnarray}
in the parametric form.

The time-dependent quasi-Heisenberg operators act in a space of
solutions of the Wheeler-DeWitt equation with the norm (12).  In the
momentum representation the formula for the average value of the
operators [13, 14] looks like the ordinary quantum-mechanics one:
\begin{equation}
\label{eq11} <A(t)>=\int \!{a}^{i\vert
k\vert}c^*(k)A(k,a,t){a}^{-i |k| }{c}(k)\biggl|_{a\rightarrow 0
}{dk}.
\end{equation}

Using (16), (17), (18) one can calculate the average value of the
scale factor of the Universe and its dispersion, for example, for
the Universe state, which is described by the wave packet
${c(k)={k}^{2}{e^{-{k}^{2}}}}$. For this aim one has to solve Eqs.
(16) and (17) numerically to find $a(k,t)$, and then substitutes it
or its square to (18).

As it is shown in the Fig. 1, the average value of the scale factor
decreases after expansion. But it does not turn to zero. The
following oscillation has smaller amplitude and finally the
evolution of the scale factor stops. This may be really interpreted
as the time disappearance. It is interesting to note that the final
static value is close to that from the static Shr\"{o}dinger's
picture discussed in the beginning of the article (the same wave
packet is used in the calculations).

Evolution of the Universe stops at the Planck times in our model.
The expansion rate may increase if one takes a wave packet with a
larger kinetic energy or a non-zero scalar field potential, which
causes the inflation of Universe. According to Ref. [6] there are
many Universes, in which the initial values of the scalar field have
some distribution. Universes, in which the value of the scalar field
is enough to produce the inflation, evolve like our Universe, while
Universes, in which the value of the scalar field is low begin to
collapse. Our paper describes namely such Universes, which come to
the static state (that is to the disappearance of time) after the
few oscillations.

To describe the real Universe, one must take to account  the
potential of the scalar field (it equals to zero in this paper). The
solution of the operator equations of motion will be much more
complicated. After all, the fundamental problem remains. According
to Ref. [13], at some time the process of self-measurement takes a
place. In this process the value of the scale factor is projected to
different values at the different areas of space. The further
evolution of these areas is classical. The modern Universe
represents one of these areas. The problem is that the homogeneous
model is insufficient to describe the process of self-measurement.

It should be noted that a time arrow appears\footnote{Certainly, the
time arrow appears also in the statistical physics and non-linear
dynamics, where some class of initial conditions leads to some
typical asymptotic in time.} in the quasi-Heisenberg picture of the
closed Universe (see also discussion in Ref. [15]). This arrow has a
direction to the end of the Universe evolution, i.e. to the static
world appearance. The arrow of time results from the unification of
gravity and quantum mechanics. There is no arrow of time in the
evolution of the classical closed Universe, because the oscillations
are symmetric relatively to the inversion of time. There is no arrow
of time in quantum mechanics either, because it does not change
after the inversion and complex conjugation  of the
Shr\mbox{\"{o}}dinger equation.

\bigskip

\small \par \noindent
[1] M. Kaku, Introduction to superstrings,
Berlin: Springer-Verlag, 1988.
\par \noindent
[2] J. A. Wheeler 1968 in: Battelle Recontres eds. B. DeWitt and
J. A. Wheeler New York, 1968,p.
\par \noindent
[3] B. S.DeWitt // Phys. Rev. 1967. Vol. 160. P. 1113.
\par \noindent
[4] A. Mostafazadeh // Annals Phys. 2004. Vol. 309 P.1; arXiv:
gr-qc/0306003.
\par \noindent
[5] E. Guendelman and A. Kaganovich // Int. J.Mod. Phys. 1993.
Vol. D 2. P. 221 (1993);  arXiv: gr-qc/0302063.
\par \noindent
[6] A.D. Linde. Particle physics and inflationary cosmology.
Harwood academic press, Switzerland (1990).
\par \noindent
[7] A. Vilenkin,  Phys. Rev. D {\bf 39}, 1116 (1989).
\par \noindent
[8] B. L. Altshuler and A. O. Barvinsky // Uspekhi. Fiz. Nauk 1996
Vol. 166 P.459 [Sov. Phys. Usp. Vol. 39 P. 429] [10] A. Vilenkin
//Phys. Rev. 1989. V. 39, P. 1116.
\par \noindent
[9] C. J. Isham, arXiv: gr-qc/9210011;
\par \noindent
[10] T. P. Shestakova and C. Simeone// Grav. Cosmol. 2004, Vol. 10
P. 161.; Ibid. Vol.10 P. 257.
\par \noindent
[11] L.D.Landau, E.M.Lifshitz, The classical Theory of Fields,
Oxford, Pergamon Press, 1982.
\par \noindent
[12] A.D. Polyanin. Handbook of exact solutions for ordianary
differential equations. New York, CRC Press,2000.
\par \noindent
[13] S.L. Cherkas, V.L. Kalashnikov, preprint.
arXiv:gr-qc/0512107.
\par \noindent
[14] S.L. Cherkas, V.L. Kalashnikov// Grav. Cosmol. 2006. V.12 P.
126.
\par \noindent
[15] C. Kiefer, H.D. Zeh//Phys. Rev. D 1995. V. 51, P. 4145.
\end{document}